\documentclass[12pt]{article}
\usepackage{epsf}

\makeatletter
\@addtoreset{equation}{section}

\renewcommand{\thefootnote}{\fnsymbol{footnote}}
\makeatother

\newcommand\be{\begin{equation}}
\newcommand\ee{\end{equation}}
\newcommand\bea{\begin{eqnarray}}
\newcommand\eea{\end{eqnarray}}

\begin{document}

\thispagestyle{empty}
\begin{flushright}
TIT/HEP--475 \\
{\tt hep-th/0203142} \\
March, 2002 \\
\end{flushright}
\vspace{3mm}
\begin{center}
{\Large
{\bf  Winding number and non-BPS bound states of walls 
in nonlinear sigma models
 }} 
\vskip 1.5cm

{\bf 
Norisuke Sakai}
\footnote{
{\it  e-mail address: nsakai@th.phys.titech.ac.jp} 
}
~and~~ {\bf 
Ryo Sugisaka}
\footnote{\it  e-mail address: sugisaka@th.phys.titech.ac.jp}

\vskip 1.5em

{ \it   
Department of Physics, Tokyo Institute of Technology 
\\
Tokyo 152-8551, JAPAN  }
\vspace{10mm}
{\bf Abstract}\\[5mm]
{\parbox{14cm}{\hspace{5mm}
Non-supersymmetric multi-wall configurations are 
generically unstable. 
It is proposed that the stabilization in compact 
space can be achieved 
by introducing a winding number into the model. 
A BPS-like bound is studied for the energy of 
configuration with nonvanishing winding number. 
Winding number is implemented in an ${\cal N}=1$ 
supersymmetric nonlinear sigma model with two chiral 
scalar fields and a bound states of BPS and anti-BPS 
walls is found to exist in noncompact spaces. 
Even in compactified space $S^1$, 
this nontrivial bound state persists above a critical 
radius of the compact dimension. 
}}
\end{center}
\vfill
\newpage
\setcounter{page}{1}
\setcounter{footnote}{0}
\renewcommand{\thefootnote}{\arabic{footnote}}

\section{Introduction}\label{INTRO}

Extended objects like walls have attracted much attention 
recently, mainly because of the possibility of ``Brane World" 
scenario where  our four-dimensional spacetime is realized on 
a wall embedded in a higher dimensional spacetime \cite{LED,RS}. 
Supersymmetry (SUSY) is one of the most promising ideas to 
solve the hierarchy problem in unified theories \cite{DGSW}. 
Walls preserving half of the original SUSY \cite{CGR,DW,KSS} are 
called $1/2$ BPS states \cite{WittenOlive}. Junctions preserving 
 $1/4$ of SUSY has also been constructed \cite{HKMN}. 
 An interesting model with two chiral scalar fields has also been 
 found allowing BPS two-wall 
 configurations \cite{SV} whose properties are studied with certain 
 Ansatz \cite{TV}. 
 By combining the brane-world scenario with SUSY, we have previously 
proposed a simple mechanism of SUSY breaking due to the coexistence 
of BPS and anti-BPS walls  \cite{MSSS}. 
We have also invented another model which allows a non-BPS 
configuration absolutely stable because of the winding number 
\cite{MSSS2}.

Motivated by  SUSY field theories in 
spacetime with dimensions higher than four 
 \cite{AlvarezFreedman}--\cite{GPT}, 
walls and junctions are studied in nonlinear sigma models 
with four supercharges \cite{NNS}. 
In order to preserve ${\cal N}=1$ SUSY in four dimensions, 
only holomorphic field redefinitions are allowed. 
With the holomorphic field redefinitions, one can transform 
SUSY field theories with minimal kinetic terms (linear sigma models) 
into those with nonminimal kinetic terms (nonlinear sigma models). 
The transformations give a model equivalent to the original 
model as far as local properties in target space are relevant. 
However, new physical effects can arise if global 
properties in target space are different. 
Typical global property of that kind is the winding number 
in target space \cite{MSSS2},  \cite{NNS}. 
BPS walls in compactified base space has been considered, 
especially in the context of two dimensional SUSY theories, 
and the importance of winding number has also been noticed 
previously \cite{HLS}, \cite{HtV}. 

A typical model admitting winding number 
is the sine-Gordon model
\begin{equation}
{\cal L} 
=-{1 \over 2} \left(\partial_\mu \psi\right)^2 
-{1 \over 2} \cos^2 \psi. 
\end{equation}
Because of the periodic field variable $\psi = \psi + 2\pi$, 
the topology of field space is $S^1$. 
When the base space is compactified ($y=y+2\pi R$) 
to $S^1 \times {\bf R}^3$, 
we obtain winding number $\pi_1(S^1)$ of the mapping 
$y \rightarrow \psi$. 
If we rewrite the same sine-Gordon model in terms of 
nonperiodic varialbe $\phi = \sin \psi$ 
\begin{equation}
{\cal L} 
=-{1 \over 2} 
{\left(\partial_\mu \phi\right)^2 \over 1-\phi^2} 
-{1 \over 2} \left(1-\phi^2\right), 
\end{equation}
it is difficult to recognize the winding number, although 
the model should be the same as long as the global property 
like winding number is irrelevant. 
For instance spectrum of fluctuations is identical for 
the zero winding number sector. 
A similar phenomenon occurs in the case of SUSY field theories. 
If we supersymmetrize the sine-Gordon model in four dimensions, 
we need a complex scalar field $\psi=\psi_R+i \psi_I$. 
The bosonic part of the Lagrangian reads 
\begin{eqnarray}
 {\cal L}
 &\!\!\!
 =
 &\!\!\!
-\partial_\mu \psi^* \partial^\mu \psi 
- \left| \cos \psi \right|^2 
 \\
 &\!\!\!
 =
 &\!\!\!
-\left(\partial_\mu \psi_R \right)^2 
-\left(\partial_\mu \psi_I \right)^2 
- \left( \cos \psi_R \cosh \psi_I\right)^2 
- \left( \sin \psi_R \sinh \psi_I\right)^2. 
    \nonumber
\end{eqnarray}
Since the real part is a periodic variable $\psi_R=\psi_R+2\pi $, 
the topology of field space is now naturally identified as 
$S^1\times {\bf R}$. 
Therefore we can define the winding number $\pi_1(S^1)$ 
of the mapping $y\rightarrow \psi_R$ from the compact base space 
$y=y+2\pi R$. 
We have previously found an exact non-BPS solution of 
two walls whose stability is guaranteed by the winding number 
for this model \cite{MSSS2}.

The purpose of this paper is to propose a general method 
to construct non-BPS configurations by introducing 
winding number and to study the properties of such 
non-BPS wall configurations, especially possible non-BPS 
bound state of walls. 
We introduce winding number by constructing a nonlinear 
sigma model. 
This can be achieved by a holomorphic field redefinition 
transforming the field variable into an angular variable 
winding around the target space of nontrivial topology 
such as $S^1$. 
We obtain a non-BPS configurations consisting of a BPS 
and an anti-BPS configuration by giving half winding 
number to each (anti-)BPS configuration. 
{}For models with single chiral scalar field with only 
real parameters, we can establish a BPS-like bound: 
configurations with nonvanishing winding number consisting 
of $n$ (anti-)BPS states has energies larger than or 
equal to the sum of energies of these $n$ (anti-)BPS states. 
Since a superposition of these $n$ (anti-)BPS states 
becomes a solution when these BPS states are far apart, 
our bound implies that no stable bound state can be 
formed in this class of models with single chiral 
scalar field. 

Although we do not find exact solutions with 
nonvanishing winding number, we can still construct 
an Ansatz of a non-BPS configuration, which is a 
superposition of BPS and anti-BPS solutions in terms 
of the periodic variable, to give nonvanishing winding number. 
 This Ansatz is tested in a model with single chiral 
scalar field and gives a repulsion between a BPS 
and anti-BPS walls and produces no bound state 
 in accordance with our BPS-like bound. 
In contrast, a similar Ansatz for configurations 
without winding number gives an attraction and shows 
annihilation into the ordinary vacuum. 
 
The model with two chiral scalar fields admits BPS 
two-wall configurations with a moduli parameter 
corresponding to the separation 
of two walls within the BPS state \cite{SV}. 
This internal structure of the BPS state offers a 
new possibility to form a bound state of BPS and 
anti-BPS state, whose stability 
is guaranteed by the nonvanishing winding number. 
We construct an Ansatz of four walls comprising 
BPS two walls and another anti-BPS two walls 
by superposing these solutions in terms of the 
periodic variable.  
We find that the BPS-like bound allows a possibility 
of configurations whose energy is lower than the 
sum of BPS and anti-BPS states. 
We evaluate the energy density of the configuration 
as a function of the moduli and of the distance 
between BPS and anti-BPS states.
We find an interesting nontrivial behavior of 
the energy density. 
We first study configurations in noncompact space 
in order to find a bound state of BPS and anti-BPS states. 
{}For one choice of intermediate vacuum, we find 
an absolute minimum of energy which is lower than 
the sum of the BPS and anti-BPS states. 
Although we use a variational Ansatz which is guaranteed 
to be a solution only in the limit of infinite separation, 
mere existence of the configurations whose energies 
are lower than the sum of the BPS and 
anti-BPS states is sufficient to conclude the 
existence of the bound state. 
The distance between BPS and anti-BPS states 
and the moduli of these states 
are approximately evaluated using our Ansatz. 
{}For another choice of intermediate vacuum, 
we find a local minimum at vanishing separation 
between BPS and anti-BPS states. 
This suggests an unstable bound state at the 
coincident limit of BPS and anti-BPS states. 
{}For compact space, we always find a minimum 
of energy when the BPS and anti-BPS states are 
equally spaced. 
This is due to a tendency to repel each other 
as indicated by the BPS-like bound. 
By the same reason, we can expect that the bound state 
that we find in the other choice of intermediate 
vacuum may disappear when the radius of the 
compact dimension is too small. 
In fact we find that the absolute minimum is 
gradually raised as the radius decreases, and 
disappears below a critical radius. 

In the next section, a method is given to introduce 
winding number by a holomorphic 
field redefinition and a BPS-like bound is derived 
for models with single chiral scalar field. 
In section 3, winding number is introduced into 
a model with two chiral scalar fields. The energy of 
non-BPS multi-wall configuration is studied numerically 
and a bound state of BPS and anti-BPS states is obtained. 

\section{Winding number in SUSY nonlinear sigma models 
} 
\label{BPS-winding}
\subsection{Introducing the winding number}
\label{sc:winding}

In order to illustrate our ideas in a simple context, 
 we consider three dimensional domain walls in four-dimensional 
 field theories with four supercharges. 
A general Wess-Zumino model with 
an arbitrary number of chiral superfields 
$\Phi^i$, 
a superpotential ${\cal W}$ 
and a K\"ahler potential $K(\Phi^i, \Phi^{* j})$ 
is given by 
\begin{equation}
{\cal L}=\left. K(\Phi^i,\Phi^{* j})  \right|_{\theta^{2}\bar{\theta}^{2}}
 + \left[\left. {\cal W}(\Phi^i) \right|_{\theta^{2}}
+ \mbox{h.c.} \right]. 
\label{generalWZmodel}
\end{equation}
We shall denote the scalar component of the superfield 
$\Phi^i(x, \theta, \bar\theta)$ as $\phi^i(x)$. 
Let us suppose that we have a wall configuration which depends only 
on $x^2=y$. 
If the following BPS equation is satisfied, two out of the 
four supercharges are conserved \cite{CGR, DW, KSS}
\begin{equation}
{\partial \phi^i \over \partial y} 
=K^{ i j^*} \frac{\partial {\cal W}^*(\phi^*)}{ \partial \phi^{*j}}.
\label{eq:BPSeq}
\end{equation}
We call such a configuration a BPS wall. 
The other two supercharges are conserved if the similar equation 
with opposite sign is satisfied 
\begin{equation}
{\partial \phi^i \over \partial y} 
=-K^{ i j^*} \frac{\partial {\cal W}^*(\phi^*)}{ \partial \phi^{*j}}.
\label{eq:antiBPSeq}
\end{equation}
We call such a configuration as an anti-BPS wall. 
Since these walls connect two supersymmetric vacua, we need models 
with two vacua at least. 
Simplest model has a single chiral scalar field $\Phi$ with a minimal 
kinetic term and a cubic superpotential ${\cal  W}$ 
\begin{eqnarray}
 {\cal L}
 &\!\!\!
 =
 &\!\!\!
 \left.\Phi^{\dagger}\Phi \right|_{\theta^{2}\bar{\theta}^{2}}
  +\left[\left.\left({m^{2}\over g}\Phi-\frac{g}{3}\Phi^{3}\right)
  \right|_{\theta^{2}}
  +{\rm h.c.}\right]
    \nonumber \\
 &\!\!\!
 =
 &\!\!\!
 -{\partial \phi \over \partial x_m} {\partial \phi^* \over \partial x^m} 
 -\left|{m^{2}\over g}-g\phi^2\right|^2+ {\rm fermions}. 
  \label{Logn} 
\end{eqnarray}
The BPS Eq.(\ref{eq:BPSeq}) and anti-BPS Eq.(\ref{eq:antiBPSeq}) 
have solutions 
\begin{equation}
\phi_{\rm (wall)}(y
) = {m \over g} \tanh \left(m (y-y_0)\right), 
\label{eq:BPSsol_cubic}
\end{equation}
\begin{equation}
\phi_{\rm (anti wall)}(y
) = - {m \over g} \tanh \left(m (y-\bar{y_0})\right), 
\label{eq:antiBPSsol_cubic}
\end{equation}
representing walls located at $y_0$ and $\bar{y_0}$ respectively. 
For compact space $y=y+2\pi R$, we have also found an exact solution 
of wall and anti-wall configuration which breaks supersymmetry 
completely \cite{MSSS} 
\begin{equation}
 \phi_{\rm (wall-antiwall)}(y)={m \over g}\frac{k\sqrt2}{\sqrt{1+k^2}}
 {\rm sn}\left({\sqrt2 \over \sqrt{1+k^2}}m y, k\right), 
 \label{acl}
\end{equation}
where ${\rm sn}(u,k)$ is the Jacobi's elliptic function, $0\leq k\leq 1$, 
 and  $R=\sqrt2\sqrt{1+k^2}K(k)/(\pi m)$, where $K(k)$ is the 
complete elliptic integral. 
This non-BPS solution corresponds to a wall located at $y=0$ and 
an anti-wall at $y=\pi R$. 
The small fluctuations around this background exhibit a tachyon 
corresponding to wall-antiwall annihilation instability \cite{MSSS}. 

A promising idea to stabilize the non-BPS configuration of two walls 
is to introduce a topological quantum number, typically a winding number 
into the model. 
We give a topology of $S^1$ to field space so that we can have 
a notion of winding from a compactified base space which is also $S^1$. 
To achieve that goal, we make a holomorphic redefinition of field $\phi$ 
into a periodic one $\psi$ 
\begin{equation}
\phi(x) = {m \over g} \sin \psi(x), \qquad 
\Phi(x, \theta, \bar\theta)= {m \over g} \sin \Psi(x, \theta, \bar\theta).
\label{eq:change_variable}
\end{equation}
In terms of the periodic variable $\psi$, the model 
(\ref{Logn}) becomes 
\begin{eqnarray}
 {\cal L}
 &\!\!\!
 =
 &\!\!\!
 \left.{m^2 \over g^2}\sin\Psi^{\dagger}\sin\Psi 
 \right|_{\theta^{2}\bar{\theta}^{2}}
  +\left.{m^3 \over g^2}\left(\sin \Psi -{1 \over 3} \sin^3\Psi\right)
  \right|_{\theta^{2}}+{\rm h.c.},   \nonumber \\
 &\!\!\!
 =
 &\!\!\!
 -{m^2 \over g^2}\left|\cos \psi {\partial \psi \over \partial x^m} \right|^2
  -\left|{m^{2}\over g}\cos^2 \psi\right|^2+ {\rm fermions}
  \label{eq:lag_periodic}.
\end{eqnarray}
The field space now acquires the topology of $S^1\times {\bf R}$. 
Then the SUSY vacua occurs at $\psi=\pi \left(n+{1 \over 2}\right)$ 
with the periodicity $\psi=\psi+2\pi$. 
The BPS equation (\ref{eq:BPSeq}) becomes 
\begin{equation}
{d \psi \over dy}=  \cos \psi. 
\label{eq:BPSeq_periodic}
\end{equation}
The BPS solution (\ref{eq:BPSsol_cubic}) is mapped into 
a solution of this transformed BPS Eq.(\ref{eq:BPSeq_periodic}) 
\begin{equation}
 \psi_{\rm (BPS)} 
(y)= {\rm arcsin} \left(\tanh \left(m (y-y_0) \right)\right)
\label{eq:solution_BPS_periodic}
\end{equation}
connecting the SUSY vacuum $\psi=-\pi/2$ at $y=-\infty$ to $\psi=\pi/2$ 
at $y=\infty$. 
The solution of the anti-BPS equation 
connecting the SUSY vacuum $\psi=\pi/2$ at $y=-\infty$ to $\psi=3\pi/2$ 
at $y=\infty$ is given by 
\begin{equation}
 \psi_{\rm (antiBPS)} (y)= 
  {\rm arcsin} \left( \tanh \left(m(y-\bar{y_0})\right)\right) + \pi .
\label{eq:solution_antiBPS_periodic}
\end{equation}
Since the value of the field $\psi$ at the right-end of the BPS wall 
(\ref{eq:solution_BPS_periodic}) is the same as the 
left-end of the anti-BPS wall (\ref{eq:solution_antiBPS_periodic}), 
there is a possibility to connect 
these two wall solutions located at $y_0 < \bar{y_0}$. 
Such a field configuration should have winding number one. 

In fact we have found previously 
that a similar model with the minimal kinetic term provides 
the same BPS equation (\ref{eq:BPSeq_periodic}) and 
that there is an exact solution for the non-BPS configuration 
of two walls for compactified space $y$ \cite{MSSS2}. 
The configuration was found to wind around the field 
space $\psi$ once and is topologically stable. 
\footnote{
Since the $1/2$-BPS solution is real, we can reinterpret 
 the inverse of the K\"ahler metric $K^{ij^*}$ as a part of 
 the derivative of a superpotential 
 $K^{ij*}\partial {\cal W}/\partial \phi^{*j}
 =\partial \tilde{\cal W}/\partial \phi^{*j}$, 
as noted in \cite{NNS}. 
}

In our model with the periodic variable (\ref{eq:lag_periodic}) 
we have an exact solution for compactified space 
\begin{equation}
 \psi_{\rm (wall-antiwall)}(y)={\rm arcsin} 
 \left(
{m \over g}\frac{k\sqrt2}{\sqrt{1+k^2}}
 {\rm sn}\left({\sqrt2 \over \sqrt{1+k^2}}m y, k\right)
  \right), 
 \label{eq:exact_sol_periodic}
\end{equation}
obtained by transforming the non-BPS solution (\ref{acl}) 
to our periodic variable 
$\psi$. 
Since  $0< \frac{k\sqrt2}{\sqrt{1+k^2}} < 1 $ for $0< k < 1$, 
the configuration has no winding number and represents the wall-antiwall 
configuration as in our original model without periodic variable 
\cite{MSSS}. 
One also finds that the small fluctuation around the background has 
exactly the same spectrum including the tachyon instability. 
This is consistent with the fact that ordinary vacuum is the minimum energy 
configuration in the vanishing winding number sector. 

\subsection{BPS-like bound for winding field configuration}
\label{sc:BPSbound}
\begin{figure}[t]
 \leavevmode
 \epsfysize=5.5cm
 \centerline{\epsfbox{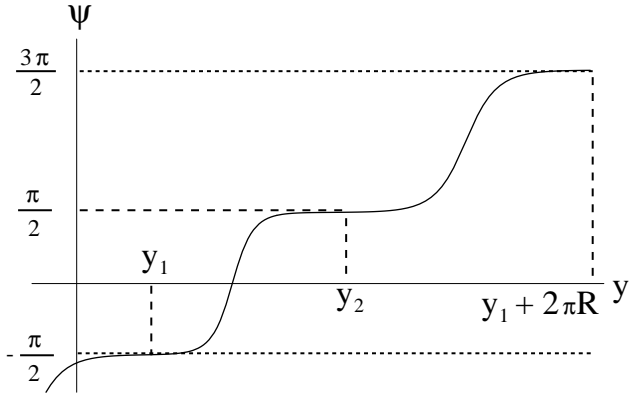}}
 \caption{The profile of the field configuration of $\psi$ with unit winding number. The dotted lines $\psi = -\pi/2$ and $\psi = 3\pi/2$ should be identified.}
 \label{psi_prof}
\end{figure}

We are interested in the field configuration with 
 a nonvanishing winding number. 
Let us consider a BPS-like bound for the energy of the configuration 
with a nonvanishing winding number. 
Let us first consider the model (\ref{eq:lag_periodic}) 
as the simplest model for illustration. 
If there is a field configuration with unit winding number, 
$\psi$ should rotate by $2\pi$ as $y$ increases by $2\pi R$. 
The noncompact space is obtained by the limit $R \rightarrow \infty$. 
Let us call the point $y_1$ where $\psi=-\pi/2$ and $y_2$ where 
$\psi=\pi/2$ and assume $y_1 < y_2$ as illustrated 
in Fig.~\ref{psi_prof}. 
The energy of a configuration with one winding number is given 
in one periodicity domain by 
\begin{eqnarray}
E 
 &\!\!\!
 =
 &\!\!\!
 \int_{y_1}^{y_1+2\pi R} dy \left[
  {m^2 \over g^2}
\left|\cos \psi {\partial \psi \over \partial y} \right|^2
 + \left|{m^{2}\over g}\cos^2 \psi\right|^2
  \right]   \nonumber \\
 &\!\!\!
 =
 &\!\!\!
 \int_{y_1}^{y_2} dy 
  \left[
\left|{m \over g}\cos \psi {\partial \psi \over \partial y} 
- {m^2 \over g}\cos^2 \psi\right|^2
+ {m^{3}\over g^2}{d \over dy}\left(\sin \psi-{1 \over 3}\sin^3 \psi\right)
  \right]  \nonumber \\
 &\!\!\!
+
 &\!\!\!
  \int_{y_2}^{y_1+2\pi R} dy \left[
\left|{m \over g}\cos \psi {\partial \psi \over \partial y} 
+ {m^2 \over g}\cos^2 \psi\right|^2
- {m^{3}\over g^2}{d \over dy}\left(\sin \psi-{1 \over 3}\sin^3 \psi\right)
  \right]   \nonumber \\
 &\!\!\!
 \geq
 &\!\!\!
 \left[{m^{3}\over g^2}\left(\sin \psi-{1 \over 3}\sin^3 \psi\right)
 \right]_{y_1}^{y_2}
- \left[{m^{3}\over g^2}\left(\sin \psi-{1 \over 3}\sin^3 \psi\right)
 \right]_{y_2}^{y_1+2\pi R}
= 2 E_{\rm BPS},
     \label{eq:energy_periodic} 
\end{eqnarray}
where $E_{\rm BPS}$ is the energy of the single BPS or anti-BPS wall. 
Therefore any configuration with unit winding number has energy larger 
than or equal to the sum of energies of a BPS wall and an anti-BPS wall. 
Since this superposition of the BPS and anti-BPS states becomes a solution 
of equation of motion as their separation goes to infinity, 
 we find that BPS state and anti-BPS state in unit winding number 
sector always repel each other as they are sufficiently far apart at least. 
Whether there is any local minimum for 
finite separation or not 
is the remaining question which we will address in the next subsection. 

This BPS-like bound can also be generalized to other models of 
a single chiral scalar field using arbitrary superpotential 
with real parameters. 
This may be achieved if the parameters of 
the model can be made real by phase rotations of the fields. 
Then we can assume that the field configuration is real. 
Let us suppose that there are two vacua at $\psi_1$ and $\psi_2$ of the 
periodic variable $\psi = \psi + 2 \pi$. 
Without loss of generality we can assume 
${\cal  W}(\psi_1) < {\cal  W}(\psi_2)$. 
If there is field configuration with a single winding number 
which takes value $\psi_1$ at $y_1$, and $\psi_2$ at $y_2$, 
we can apply a BPS bound for the interval $y_1< y < y_2$ 
and anti-BPS bound for the interval $y_2 < y < y_1 + 2\pi R$ 
as in Eq.(\ref{eq:energy_periodic}). 
Thus we obtain the energy of the field configuration with single winding 
number is bounded by 
$2E_{\rm BPS}=2\left({\cal  W}(\psi_2)-{\cal  W}(\psi_1)\right)$. 
Similarly we can easily find that winding field configurations 
consisting of $n$ (anti-)BPS states in one periodicity domain 
$0 \le y \le 2\pi R$ 
have energy larger than or equal to the sum of these BPS states. 

\subsection{Repulsion between BPS and anti-BPS walls}
\label{sc:BPS_antiBPS_potential}
Since we cannot find exact solutions in the sector with nonvanishing 
winding number, we shall use approximate evaluation of the possible 
configurations inspired by the BPS (\ref{eq:solution_BPS_periodic}) 
and 
anti-BPS (\ref{eq:solution_antiBPS_periodic}) solutions. 
This is at least sufficient to give an upper bound of the energy of the 
possible minimum energy solution from the viewpoint of 
a variational approach. 

\begin{figure}[h]
 \leavevmode
 \epsfysize=4.5cm
 \centerline{\epsfbox{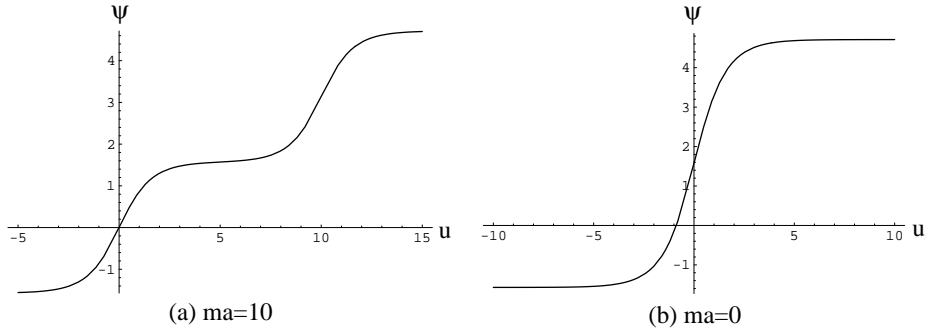}}
 \caption{The profile of $\psi$ with unit winding number, 
 the superposition of the BPS wall and the anti-BPS wall 
 at $ma=10$ (a) and $ma=0$ (b).}
 \label{w1prof}
\end{figure}

Let us consider a superposition of the BPS wall solution 
(\ref{eq:solution_BPS_periodic}) centered at $y=0$ 
and the anti-BPS solution (\ref{eq:solution_antiBPS_periodic}) 
centered at $y=a$ 
\begin{equation}
\psi(y)= 
 {\rm arcsin} \left(\tanh \left(m y \right)\right)
 + 
 {\rm arcsin} \left(\tanh \left(m (y-a) \right)\right) + {\pi \over 2}
\label{eq:2WALL}
\end{equation}
connecting the SUSY vacuum $\psi=-\pi/2$ at $y=-\infty$ to $\psi=3\pi/2$ 
at $y=\infty$, and has unit winding number. 
Although this is not a static solution of the equation of motion for 
finite separation $a$, 
it reduces to a solution in the limit $a \rightarrow \infty$. 
Defining a dimensionless coordinate 
\begin{equation}
u \equiv my, \qquad u_a \equiv ma, 
\end{equation}
the energy of this configuration is found to be 
\begin{eqnarray}
E 
 &\!\!\!
 =
 &\!\!\!
 \int_{-\infty}^{\infty} dy \left[
  {m^2 \over g^2}
\left|\cos \psi {\partial \psi \over \partial y} \right|^2
 + \left|{m^{2}\over g}\cos^2 \psi\right|^2
  \right]   \nonumber \\
 &\!\!\!
 =
 &\!\!\!
{m^3 \over g^2} \int_{-\infty}^{\infty} du 
\left[  \left({{\rm tanh}u \over  {\rm cosh}(u-u_a)}+ 
  {{\rm tanh}(u-u_a) \over {\rm cosh}u}\right)^2
\left({1 \over {\rm cosh}u}+{1 \over {\rm cosh}(u-u_a)} \right)^2
\right.
  \nonumber \\
 &\!\!\!
 &\!\!\! 
 +
 \left. \left({{\rm tanh}u \over  {\rm cosh}(u-u_a)}+
  {{\rm tanh}(u-u_a) \over {\rm cosh}u}\right)^4
\right].
     \label{eq:energy_2wall} 
\end{eqnarray}
One should note that the field (\ref{eq:2WALL}) as well 
as its derivative are non-singular in the entire 
region of $y$. 
On the other hand, the energy density in the integrand of 
Eq.(\ref{eq:energy_2wall}) has 
contributions from the kinetic term (first term) 
and the potential term (second term), 
both of which have powers of $\cos \psi$ 
vanishing at vacua. 
As a consequence, the energy density of the two wall 
configuration in Eq.(\ref{eq:energy_2wall}) 
has a zero at $y=a/2$ for any values of $a \ge 0$ 
and exhibits two separated peaks for two walls. 
This is true even for $a = 0$ 
where the field $\psi$ itself shows no sharp separation of 
two walls as shown in Fig.\ref{w1prof}b. 

\begin{figure}[h]
 \leavevmode
 \epsfysize=4.5cm
 \centerline{\epsfbox{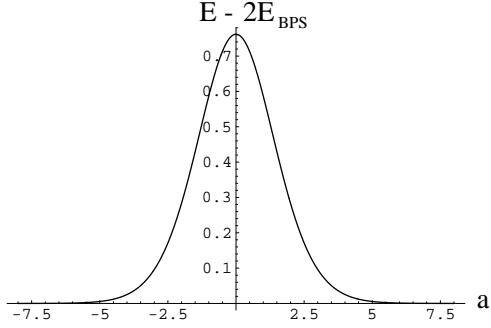}}
 \caption{The energy of $\psi$ as a function of the wall separation $a$ 
 ($m=1, g=1$). }
 \label{w1ve}
\end{figure}

A typical field configuration at $u_a = ma =10 $ in Fig.~\ref{w1prof}a
shows $\psi$ winding once to form two walls. 
Even at the coincident limit $a \rightarrow 0$ of two walls, 
 field configuration is nontrivial as shown in Fig.~\ref{w1prof}b. 
The energy as a function of the wall separation $a$ is shown in 
{}Fig.~\ref{w1ve}, 
where the parameters are set to $m=1, \; g=1$. 
We see that the energy is always larger than the sum of the isolated 
wall and anti-wall and reduces to the sum at $a \rightarrow \infty$ 
in accordance 
with our BPS-like bound. 
Therefore we find that BPS wall and anti-BPS wall repel each other 
and have no stable bound state in the unit winding number 
sector. 

To examine how well our Ansatz carries the correct behavior of the 
lowest energy configuration, we also compute the energy of the 
corresponding Ansatz in vanishing winding number sector: 
\begin{equation}
\psi_{\rm no \: winding}(y)= 
 {\rm arcsin} \left(\tanh \left(m y \right)\right)
 - 
 {\rm arcsin} \left(\tanh \left(m (y-a) \right)\right) - {\pi \over 2}. 
\label{eq:WALLantiWALL}
\end{equation}
\begin{eqnarray}
E_{\rm no \: winding} 
 &\!\!\!
 =
 &\!\!\!
{m^3 \over g^2} \int_{-\infty}^{\infty} du 
\left[  \left({{\rm tanh}u \over {\rm cosh}(u-u_a)} - 
  {{\rm tanh}(u-u_a) \over {\rm cosh}u}\right)^2
\left({1 \over {\rm cosh}u}-{1 \over {\rm cosh}(u-u_a)} \right)^2
\right.
  \nonumber \\
 &\!\!\!
 &\!\!\! 
 +
 \left. \left({{\rm tanh}u \over {\rm cosh}(u-u_a)} - 
  {{\rm tanh}(u-u_a) \over {\rm cosh}u}\right)^4
\right]. 
     \label{eq:energy_wall_antiwall} 
\end{eqnarray}
\begin{figure}[b]
 \leavevmode
 \epsfysize=4.5cm
 \centerline{\epsfbox{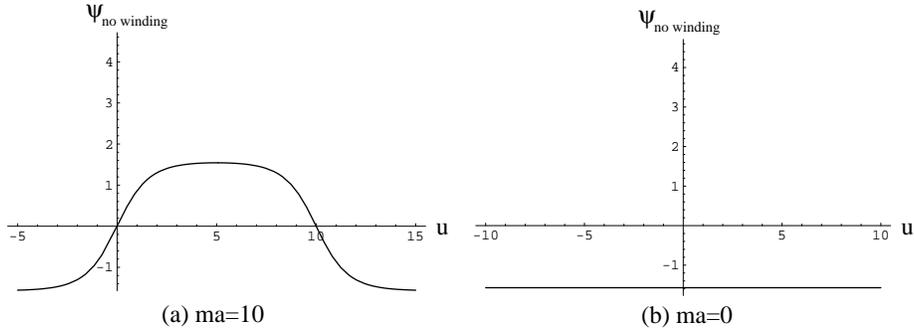}}
 \caption{The profile of $\psi_{\rm no\; winding}$ without winding number, 
 the superposition of the BPS wall and the anti-BPS wall 
 at $ma=10$ (a) and $ma=0$ (b).}
 \label{w0prof}
\end{figure}
\begin{figure}[h]
 \leavevmode
 \epsfysize=4.5cm
 \centerline{\epsfbox{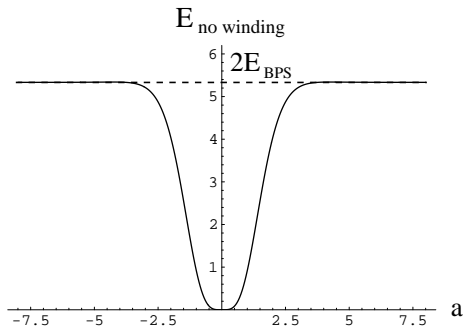}}
 \caption{The energy of $\psi_{\rm no\; winding}$ as a function of 
 the wall separation $a$ ($m=1, g=1$).  }
 \label{w0ve}
\end{figure}
A typical field configuration at $u_a=10$ in Fig.~\ref{w0prof}a 
shows no winding to be compared with Fig.~\ref{w1prof}a. 
At the coincident limit $a \rightarrow 0$ of two walls, the field $\psi$ 
becomes constant as shown in Fig.~\ref{w0prof}b and reduces to the ordinary 
vacuum $\psi=-\pi/2$ in contrast to the unit winding number 
case in Fig.~\ref{w1prof}b. 
In Fig.~\ref{w0ve} we show the energy of 
the two wall configuration in the vanishing winding number sector 
as a function of the wall separation $a$. 
It reduces to the sum of the BPS energies of two walls at 
$a \rightarrow \pm\infty$ and vanishes at the coincident point $a=0$. 
This clearly shows that the wall-antiwall configuration in the zero 
winding number sector is unstable and annihilates into the vacuum. 

\section{Winding number in a model with two fields 
} 
\label{sc:wind_two_fields}

\subsection{Model with two fields}
\label{sc:two_fields_model}

In order to explore a nontrivial behavior of winding number configuration, 
we consider the next simplest possibility, 
models with two chiral scalar fields. 
It has been found that the model with minimal kinetic terms for chiral 
scalar fields $\Phi$ and $X$ with the following superpotential ${\cal  W}$ 
has an integral of motion  \cite{SV} 
\begin{equation}
 {\cal  W}(\Phi,X)
=\frac{m^{2}}{g}\Phi-\frac{g}{3}\Phi^{3}-{g \over 4}\Phi X^{2}, 
\quad m, g > 0. 
\end{equation}
This  model is the simplest modification of our model in Eq.(\ref{Logn}) 
in the previous section to allow 
four degenerate SUSY vacua at 
$(\phi, \chi)=(\pm{m \over g}, 0), (0, \pm 2{m \over g})$. 
There are BPS wall solutions connecting the vacuum 
$(-{m \over g}, 0)$ to $(0, \pm 2{m \over g})$ and solutions 
$(0, \pm 2{m \over g})$ to $({m \over g}, 0)$. 
Since both of them turn out to conserve the same supercharge, 
the smooth connection of these wall solutions located far apart 
should be a solution of the same BPS equation. 
A remarkable property of this model is that it admits a BPS solution of 
two walls connecting 
$(\phi, \chi)=(-{m \over g}, 0)$ to $({m \over g}, 0)$ 
\begin{eqnarray}
 \phi
 &\!\!\!
=
 &\!\!\!
\frac{m}{g}f(u), 
 \qquad f(u)={ \sinh u \over \cosh u + t},  \nonumber \\
 \chi
 &\!\!\!
=
 &\!\!\!
 \pm \frac{m}{g}h(u), 
 \qquad 
h(u)= 2\sqrt{t \over \cosh u + t}, 
\label{eq:BPS_2_wall}
\end{eqnarray}
where $0 \le t $ is a moduli parameter \cite{SV, TV}. 
This configuration can be interpreted as a smooth connection of the above two 
BPS walls connecting between 
$(-{m \over g}, 0)$ and $(0, \pm 2{m \over g})$ and between 
$(0, \pm 2{m \over g})$ to $({m \over g}, 0)$. 
They are centered at $y=0$. If the moduli parameter $t$ is larger than one, 
these two walls are separated by a distance $y=\hat t$ where 
${\rm cosh} (m\hat t)=t$. 
The case with $0 \le t < 1$ corresponds to two walls compressed each other 
so that the walls merge together completely. 

We introduce the concept of winding number by making a holomorphic change 
of variable (\ref{eq:change_variable}) from $\phi$ to 
a periodic one $\psi={\rm arcsin} (g\phi/m)$. 
The bosonic part of our model with the periodic variable is then given by 
\begin{eqnarray}
 {\cal L}_{\rm bosonic}
 &\!\!\!
 =
 &\!\!\!
 -{m^2 \over g^2}\left|\cos \psi {\partial \psi \over \partial x^m} \right|^2
 -\left|{\partial \chi \over \partial x^m} \right|^2
  -\left|{m^{2}\over g}\cos^2 \psi-{g \over 4}\chi^2\right|^2 
  - \left|-{m \over 2} \chi \sin \psi\right|^2.
  \label{eq:lag_periodic_two_fields} 
\end{eqnarray}
The vacua of this model are at 
\begin{equation}
 (\psi, \chi)=\left({\pi \over 2}+n \pi, 0\right), \qquad 
 \left(n \pi, {2 m \over g}\right), \qquad 
 \left(n \pi, -{2 m \over g}\right), 
\end{equation}
with integer $-\infty < n < \infty$. 
The BPS equations are given by 
\begin{equation}
 \cos \psi{\partial \psi \over \partial u}
 =\cos^2 \psi^* -{g^2 \over 4m^2}\chi^{*2},  \qquad 
 {\partial \chi \over \partial u} =- {1 \over 2}\sin \psi^*\chi^*, \qquad 
 u\equiv my.
\end{equation}
One can obtain from (\ref{eq:BPS_2_wall}) a BPS two-wall solution 
which connects $(\psi, \chi)=(-\pi/2, 0)$ at $y=-\infty$ to 
$(\psi, \chi)= (\pi/2, 0)$ at $y=\infty$ 
\begin{equation}
 \psi_{\rm BPS}={\rm arcsin }\left(f(u)
\right), 
 \qquad 
 \chi^{\pm}_{\rm BPS}= \pm \frac{m}{g}h(u)
,
\label{eq:BPS_2_wall_periodic}
\end{equation}
where the functions $f(u), h(u)$ are defined in (\ref{eq:BPS_2_wall}). 
The ${}^{\pm}$ of $\chi^{\pm}$ corresponds to the sign of vacuum 
 $(\psi, \chi)=(0, \pm 2m/g)$ in the intermediate region. 
Another two wall configuration connecting $(\psi, \chi)=(\pi/2, 0)$ 
at $y=-\infty$ to $(\psi, \chi)= (3\pi/2, 0)$ at $y=\infty$ is 
given as a solution of anti-BPS equation preserving the opposite 
combination of supercharges 
\begin{equation}
 \psi_{\rm antiBPS}
 ={\rm arcsin }\left(f(u)
\right) + \pi, 
 \qquad 
 \chi^{\pm}_{\rm antiBPS}= \pm \frac{m}{g}h(u)
. 
\label{eq:antiBPS_2_wall_periodic2}
\end{equation}
These solutions are centered at $y=0$ and have a moduli $t$. 

\begin{figure}[h]
 \leavevmode
 \epsfysize=7cm
 \centerline{\epsfbox{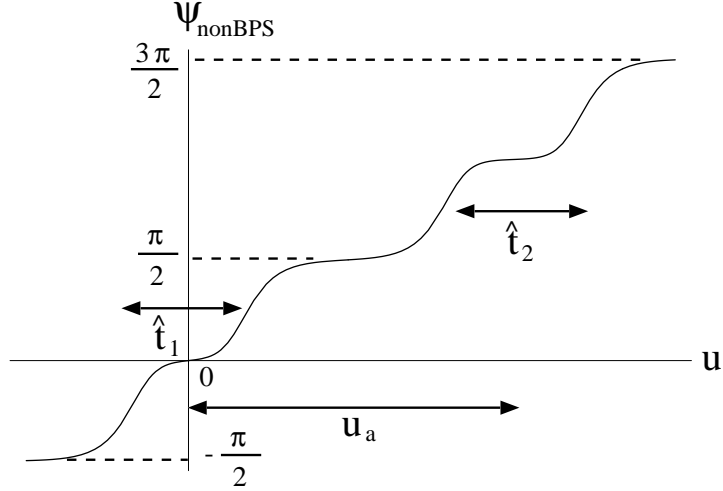}}
 \caption{The profile of 
 the field configuration of $\psi_{\rm nonBPS}$.  }
 \label{2f_phi_prof}
 \end{figure}
 \begin{figure}[h]
 \leavevmode
  \epsfysize=4.5cm
 \centerline{\epsfbox{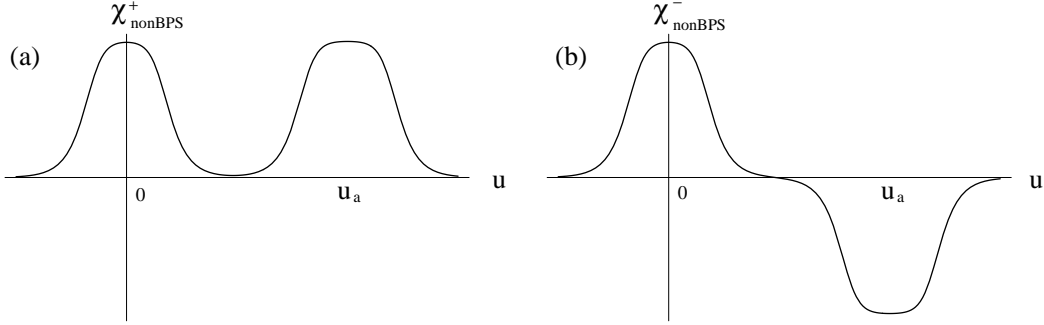}}
 \caption{The profile of the field configuration of 
 $\chi^\pm_{\rm nonBPS}$.}
 \label{2f_chi_prof}
\end{figure}

We can construct an Ansatz for the configuration with unit 
winding number by superposing the BPS two-wall solution with 
the moduli $t_1$ centered at $y=0$ and the anti-BPS two wall solution 
with the moduli $t_2$ centered at $y=a$ 
\begin{eqnarray}
 \psi_{\rm nonBPS}
 &\!\!\!
 =
 &\!\!\!
{\rm arcsin }\left({ \sinh u \over \cosh u + t_1}\right) +  
{\rm arcsin }\left({ \sinh (u-u_a) \over \cosh (u-u_a) + t_2}\right)
 + {\pi \over 2},   \nonumber \\
 \chi^{\pm}_{\rm nonBPS}
 &\!\!\!
 =
 &\!\!\!
 \frac{2m}{g}\left(\sqrt{t_1 \over \cosh u + t_1}
\pm \sqrt{t_2 \over \cosh (u-u_a) + t_2} \right),
\label{eq:4_wall_Ansatz}
\end{eqnarray}
where the sign ${}^{\pm}$ for the field $\chi^{\pm}$ indicates the 
same ($+$) or opposite ($-$) sign of vacuum in the intermediate region 
around $y=0$ and around $y=a$. 
A typical field configuration $(\psi, \chi^+)$ 
is illustrated in Figs.~\ref{2f_phi_prof} and \ref{2f_chi_prof}a. 
We have shown the distance $\hat t_i\equiv {\rm arccosh}( t_i)/m$ 
between walls within the BPS state($i=1$) and the anti-BPS state ($i=2$). 
 For comparison, field configuration  $(\psi, \chi^-)$ for the vacuum 
$(\psi, \chi)=(0, -2m/g)$ in the intermediate region 
is illustrated in Figs.~\ref{2f_phi_prof} and \ref{2f_chi_prof}b.

\subsection{BPS-like bound for two fields model}
\label{sc:bound_two_fields_model}

Let us first examine what sort of BPS-like bound can be obtained 
in the case of two fields. 
The energy in the one periodicity domain is given by 
\begin{equation}
E 
 =
 \int_{0}^{2\pi R} dy \left[ 
\left|{m \over g}\cos \psi {d \psi \over d y} \right|^2
+\left|{d \chi \over d y}  \right|^2
+\left|{m^{2}\over g}\cos^2 \psi-{g \over 4}\chi^2 \right|^2
+\left|-{m \over 2} \chi \sin \psi\right|^2\right].
   \label{eq:totalenergy_two_field} 
\end{equation}
Similarly to the single field model, we assume that the periodic 
field $\psi$ takes values $-\pi/2$ at $y_1$ and $\pi/2$ at $y_2$. 
We find 
\begin{eqnarray}
E 
 &\!\!\!
 =
 &\!\!\!
 \int_{y_1}^{y_2} dy \left[ 
\left|{m \over g}\cos \psi {d \psi \over d y} 
-{m^{2}\over g}\cos^2 \psi+{g \over 4}\chi^2 \right|^2
+\left|{d \chi \over d y} 
+{m \over 2} \chi \sin \psi\right|^2
+{d {\cal W}(\psi, \chi) \over dy}
  \right]  \\
 &\!\!\!
+
 &\!\!\!
  \int_{y_2}^{y_1+2\pi R} dy \left[
\left|{m \over g}\cos \psi {\partial \psi \over \partial y} 
+{m^{2}\over g}\cos^2 \psi-{g \over 4}\chi^2 \right|^2
+\left|{d \chi \over d y} 
-{m \over 2} \chi \sin \psi\right|^2
-{d {\cal W}(\psi, \chi) \over dy}
  \right].   \nonumber  
     \label{eq:energy_two_field} 
\end{eqnarray}
In the case of single field model, one can specify the value of the 
superpotential in terms of the value of the field $\psi$. 
Combined with the requirement of nonvanishing winding number and the 
continuity of a real function $\psi(y)$, we are 
sure that the superpotential has to reach the vacuum value 
${\cal  W}(\psi=\pi/2)$ before returning back to the original 
value ${\cal  W}(\psi=-\pi/2)$ as dictated 
by the periodicity of the superpotential in our model. 
This is the origin of the BPS-like bound for the single field model. 
In the case of two fields, however, 
the above bound reads 
\begin{eqnarray}
E 
 &\!\!\!
\ge
 &\!\!\!
 2\left[{\cal W}\left(\psi(y_2)={\pi \over 2}, \chi(y_2)\right)
-{\cal W}\left(\psi(y_1)={-\pi \over 2}, \chi(y_1)\right)\right]
 \nonumber \\
 &\!\!\!
 =
 &\!\!\!
 2\left[
 \left({2m^3 \over 3g^2} -{m \over 4}\chi^2(y_2)\right)
- \left(-{2m^3 \over 3g^2} +{m \over 4}\chi^2(y_1)\right)
\right]
 \nonumber \\
 &\!\!\!
 =
 &\!\!\!
 2E_{\rm BPS} -{m \over 2}\left(\chi^2(y_2)+\chi^2(y_1)\right).
\end{eqnarray}
Nonvanishing $\chi(y_1)$ and $\chi(y_2)$ make this bound lower than 
twice the BPS energy. 
In the case of two fields, it is not guaranteed for two fields 
to take particular value specified by the vacuum at the same time. 
Consequently the superpotential need not reach 
the value at the vacuum before returning back to its original value. 
Since we have just chosen particular points $y_1, y_2$ to divide the 
integration region to evaluate the BPS or anti-BPS bound, 
it is possible that we can have better bound by choosing other 
point of division. 
However, it does not seem to be possible to choose such a point 
in general situations. 
Therefore we just conclude that the energy of the winding configuration 
need not be larger than the sum of BPS and anti-BPS states. 
This result suggests that there is a possibility for a non-BPS bound state 
of walls in the case of the two-wall model. 

\subsection{Non-BPS multi-walls in the noncompact space}
\label{sc:noncompact_space}

Here we present energy (\ref{eq:totalenergy_two_field}) 
of the non-BPS configuration of four walls 
with unit winding number (\ref{eq:4_wall_Ansatz}). 
To avoid too many parameters, we will use the same moduli parameters 
for both BPS and anti-BPS states $t_1=t_2\equiv t$ except stated otherwise. 

\begin{figure}[b]
 \leavevmode
 \epsfysize=4.5cm
 \centerline{\epsfbox{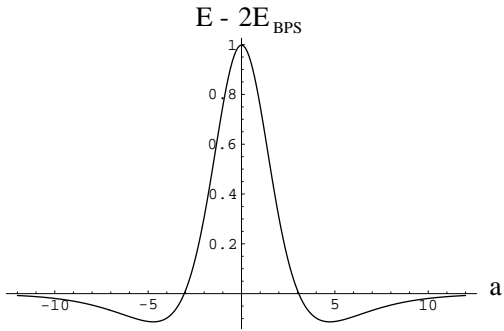}}
 \caption{The energy of the field configuration of $(\psi,\chi^+)$ 
 as a function of $a$ at $t=0.2$ ($m=1, g=1$).}
 \label{ve+nc}

\end{figure}
\begin{figure}[h]
 \leavevmode
 \epsfysize=4.5cm
 \centerline{\epsfbox{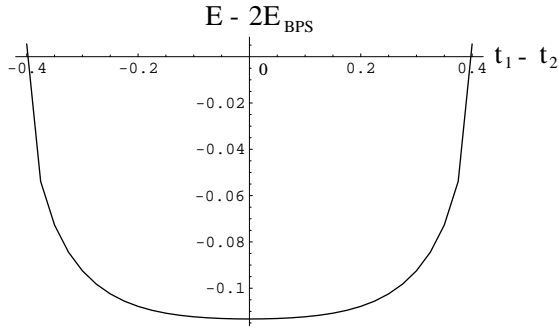}}
 \caption{The energy of $(\psi,\chi^+)$ as a function of $t_1-t_2$ 
 at $(t_1 +t_2)/2=0.2$ and $ma =4.73$ ($m=1, g=1$). }
 \label{ve+nc_t}
\end{figure}

{}For the case of $\chi^{\pm}$ with the same sign 
of the field $\chi$ for BPS and anti-BPS 
walls, we show the energy as a function of separation $a$ between 
BPS and anti-BPS walls, for fixed moduli $t$ in Fig.~\ref{ve+nc}. 
A typical field configuration can be obtained by letting $\hat t_1=\hat t_2$ 
in Figs.~\ref{2f_phi_prof} and \ref{2f_chi_prof}a. 
We observe the following:
\begin{enumerate}
\item
There exist configurations which have energies lower than the sum of BPS and 
anti-BPS states in contrast to the single field case. 
Although we have only a limited Ansatz of field configurations 
inspired by physical 
considerations, this fact clearly shows that the BPS and anti-BPS states 
infinitely far apart is not the lowest energy state in the nonvanishing 
winding number sector. 
\item
Among these configurations, we find the minimum energy configuration 
which has a separation $a \approx 4.73/m
$ of BPS and anti-BPS states 
and the moduli $t \approx 0.200$ which is shown in Fig.\ref{ve+nc}. 
Thus we find an approximate evaluation of the minimum energy configuration 
for the unit winding number sector. 
\item
We have also examined the energy as a function of the difference 
$t_1-t_2$ between 
the moduli parameters of the BPS state $t_1$ and the anti-BPS state $t_2$ 
as shown in Fig.~\ref{ve+nc_t}. 
It shows clearly the configuration achieves the minimum energy 
for $t_1=t_2$. 
\item
Therefore we find that there exists a non-BPS bound state of walls 
whose approximate configuration is given by BPS and anti-BPS two walls 
with moduli parameter $t_1=t_2 \approx 0.200
$ which are separated by 
$a \approx 4.73/m$. 
\end{enumerate}

\begin{figure}[h]
 \leavevmode
 \epsfysize=4.5cm
 \centerline{\epsfbox{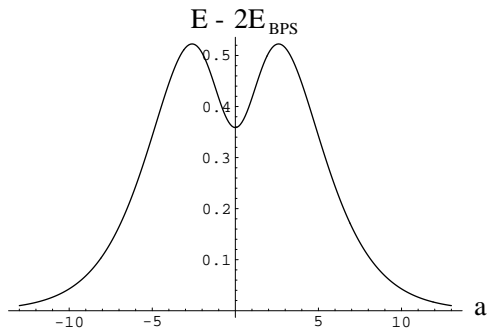}}
 \caption{The energy of the field configuration of $(\psi,\chi^-)$ as 
 a function of $a$ at $t=0.423$ ($m=1, g=1$). }
 \label{ve-nc}
\end{figure}

{}For the case with the opposite sign of the fields $\chi$ for 
BPS and anti-BPS walls, we show the energy as a function of separation 
$a$ between BPS and anti-BPS walls, for fixed moduli $t \approx 0.423$ in 
Fig.~\ref{ve-nc}. 
A typical field configuration can be obtained by letting $\hat t_1=\hat t_2$ 
in Figs.~\ref{2f_phi_prof} and \ref{2f_chi_prof}b. 
We observe the following:
\begin{enumerate}
\item
We find that energy is always higher than the sum of the BPS and anti-BPS 
states. 
\item
There exists a local minimum of energy at zero separation $a=0$ between 
BPS and anti-BPS states for the moduli parameter $t < 3.705$. 
The lowest value of this local minimum occurs at the moduli 
$t \approx 0.423
$ which is shown in Fig.~\ref{ve-nc}. 
\end{enumerate}

\subsection{Non-BPS multi-walls in compactified space}
\label{sc:compact_space}

It is interesting to examine if the above non-BPS bound state of 
walls persists when the space is compactified on $S^1$, since 
one expects a repulsion from the other walls located at 
$2\pi R$ which is the periodicity of the base space $y=y+2\pi R$. 

When the space is compactified, the BPS states 
are placed at $y=2n\pi R$ and anti-BPS walls at 
$y=2n\pi R + a$ periodically. Then we have an Ansatz 
\begin{eqnarray}
 \psi_{\rm nonBPS}
 &\!\!\!
 =
 &\!\!\!
 \sum_{n=-\infty}^{\infty}\left[
{\rm arcsin }\left(f(u-2nu_0)
\right) +  
{\rm arcsin }\left(f(u-u_a-2nu_0)
\right)
 + \pi \right] - {\pi \over 2},   \nonumber \\
 \chi^{\pm}_{\rm nonBPS}
 &\!\!\!
 =
 &\!\!\!
 \sum_{n=-\infty}^{\infty}\left[
 \frac{m}{g}\left(h(u-2nu_0)
\pm h(u-u_a-2nu_0)
\right)
\right],
\label{eq:2n_wall_Ansatz}
\end{eqnarray}
where $u\equiv m y, u_a\equiv ma$ and $u_0\equiv m\pi R$. 

\begin{table}[b]
\begin{center}
\begin{tabular}{|c||c|c|c|} \hline 
$2m\pi R$ & $10$ & $20$ & $40$ \\ \hline \hline
$a=0,2\pi R$ & $9.18 \times 10^{-5}$ &$4.12 \times 10^{-9}$ &$\ll 10^{-9}$ \\ \hline
$a=\pi R$ & $1.03 \times 10^{-2}$ &$6.81 \times 10^{-5}$ &$3.09 \times 10^{-9}$ \\ \hline
\end{tabular}
\caption{Fractional difference $(E_5-E_3)/E_3$ between energies in 
compact space in approximations with five walls $E_5$ and 
three walls $E_3$ for $2m\pi R=10, 20, 40$. 
Difference is largest for anti-wall placed at the center of periodicity 
$a=\pi R$, and smallest at coincident limit $a=0,2\pi R$. }
\label{3vs5wall}
\end{center}
\end{table}
In one periodicity domain $a-\pi R < y < a + \pi R$, only nearby walls 
are important. 
Three walls (placed at $y=0, a, 2\pi R$) and five walls 
 (placed at $y=a-2\pi R, 0, a, 2\pi R, a+2\pi R$) 
 approximations of energy in an interval 
 $a-\pi R < y < a+\pi R$ are compared in the case of 
 the moduli parameter $t=0$. 
We find an excellent agreement between three wall and five wall 
approximations. 
We show the fractional difference $(E_5-E_3)/E_3$ between energies in 
compact space in approximations with five walls $E_5$ and 
three walls $E_3$ in Table.~\ref{3vs5wall}. 
Therefore we choose to use three wall approximation in the following. 
The energy density in Eq.(\ref{eq:totalenergy_two_field}) 
can be evaluated in the three wall approximation 
by using the $f_i=f(u_i), h_i=h(u_i), i=1, 2, 3$, and 
$u_1=u, u_2=u-u_a, u_3=u-2m\pi R$ 
\begin{eqnarray}
\sin \psi 
 &\!\!\!
 =
 &\!\!\!
-f_1\sqrt{(1-f_2^2)(1-f_3^2)}
-f_2\sqrt{(1-f_3^2)(1-f_1^2)}
-f_3\sqrt{(1-f_1^2)(1-f_2^2)}
+f_1f_2f_3,
\nonumber  \\
\cos \psi 
 &\!\!\!
=
 &\!\!\!
-\sqrt{(1-f_1^2)(1-f_2^2)(1-f_3^2)}
+f_1f_2\sqrt{(1-f_3^2)}
+f_2f_3\sqrt{(1-f_1^2)}
+f_3f_1\sqrt{(1-f_2^2)},
\nonumber  \\
{d \psi \over d u}
 &\!\!\!
=
 &\!\!\!
\sum_{i=1}^3
{1 \over \sqrt{(e^{u_i}+t)(e^{-u_i}+t)}}
{t {\rm cosh} u_i + 1 \over {\rm cosh} u_i + t}, 
\qquad i=1, 2, 3.
\end{eqnarray}

\begin{figure}[t]
 \leavevmode
 \epsfysize=4.2cm
 \centerline{\epsfbox{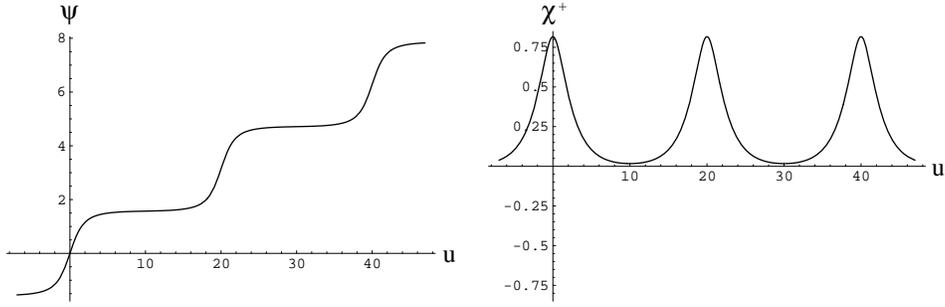}}
 \caption{The profile of the field configuration of $(\psi,\chi^+)$ in 
 three wall approximation at $2m\pi R = 40$ and $t=0.2$ ($m=1, g=1$).}
 \label{c+prof}
 \end{figure}
\begin{figure}[t]
 \leavevmode
 \epsfysize=4.2cm
 \centerline{\epsfbox{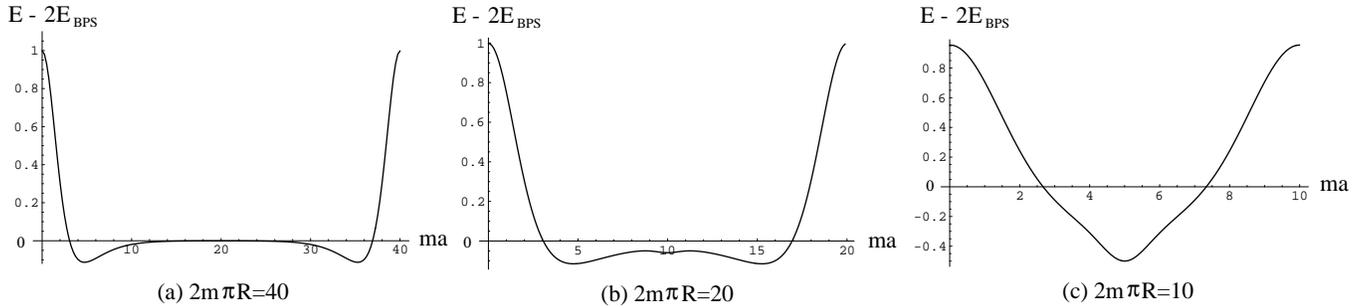}}
 \caption{The energy of  $(\psi,\chi^+)$ as a function of $ma$ at $t=0.2$ in the case of $2m\pi R =40$ (a), $2m\pi R =20$ (b) and $2m\pi R =10$ (c) ($m=1, g=1$). }
 \label{c+veR20}
\end{figure}
{}For the choice of same sign of $\chi$ for BPS and anti-BPS states, 
a typical field configuration $\psi$ and $\chi^+$ 
is shown 
in Fig.~\ref{c+prof} 
for $2m\pi R=40$ and $t=0.2$ choosing $a=\pi R$. 
The corresponding energy of the unit winding number configuration is 
shown in Fig.~\ref{c+veR20}a as a function of $ma$. 
We find that there is an absolute minimum at $ma =4.75$ which is 
identical to the noncompact case. 
We also find 
a very shallow local minimum at the center $a=\pi R$. 
This comes about because of the compactification. 
General tendency of the non-BPS walls is that they exert repulsion 
as we have encountered in the single field case. 
This repulsion produces a minimum at the central position. 

As we decrease the compactification radius $R$, the absolute minimum 
gets shallower as shown in Fig.~\ref{c+veR20}b for the case of 
$2m\pi R=20$ with the same moduli $t=0.2$. 
Eventually the absolute minimum around $ma=4.75$ disappears 
and we obtain only single minimum at the center $a=\pi R$ as shown 
in Fig.~\ref{c+veR20}c. 
Thus we find that the non-BPS bound state of walls 
persists for larger values of 
compactification radius up to $2m\pi R < 16.92$, and that it becomes 
unstable for smaller values of the radius. 

\begin{figure}[h]
 \leavevmode
 \epsfysize=4.2cm
 \centerline{\epsfbox{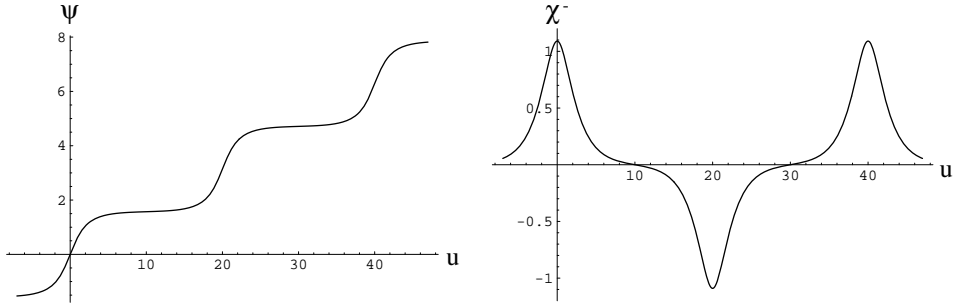}}
 \caption{The profile of the field configuration of $(\psi,\chi^-)$ in 
 three wall approximation at $2m\pi R = 40$ and $t=0.423$.}
 \label{c-prof}
 \end{figure}
\begin{figure}[h]
 \epsfysize=4.2cm
 \centerline{\epsfbox{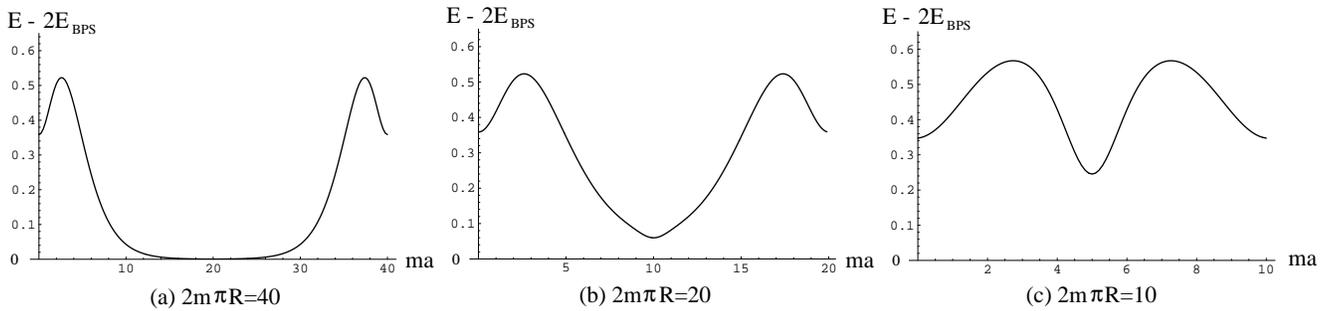}}
 \caption{The energy of  $(\psi,\chi^-)$ as a function of $ma$ 
 at $t=0.423$ in the case of $2m\pi R =40$ (a), $2m\pi R =20$ (b)
  and $2m\pi R =10$ (c) ($m=1, g=1$).}
 \label{c-veR20}
\end{figure}
{}Let us next examine the case with the opposite sign 
of the fields $\chi$ for BPS and anti-BPS walls. 
A typical field configuration $\psi$ and $\chi^-$ is shown 
in Fig.~\ref{c-prof} 
for $2m\pi R=40$ and $t=0.423$ choosing $a=\pi R$. 
{}For noncompact space, we have found a local minimum of energies 
at $a=0$. 
This still persists for $2m\pi R=40$, $2m\pi R=20$ and $2m\pi R=10$ 
as shown in Fig.~\ref{c-veR20}.
The absolute minimum always occurs at the center $a=\pi R$ 
in the case of the opposite sign 
of the fields $\chi$ for BPS and anti-BPS walls. 
Since the width of the BPS state is of order $1/m$, 
our Ansatz requires $my$ to span a region of a few times 
$1/m$ for the field $\psi$ to make a full $2\pi$ 
winding. 
Therefore we should not trust our Ansatz for too small values of 
radius $R$. 


\renewcommand{\thesubsection}{Acknowledgments}
\subsection{}

We thank Nobuhito Maru and Yutaka Sakamura for a collaboration of 
previous works where an idea of this work arises. 
This work is supported in part by Grant-in-Aid for Scientific 
Research from the Ministry of Education, Culture, Sports, 
Science and 
Technology, Japan, priority area (\#707) ``Supersymmetry and 
unified theory
of elementary particles" and No.13640269. 
R.S.~is supported 
by the Japan Society for the Promotion of Science for Young 
Scientists (No.6665).


\end{document}